\pdfoutput=1

\documentclass[11pt]{article}

\usepackage[preprint]{acl}

\usepackage{times}
\usepackage{latexsym}
\usepackage{amsmath}
\usepackage{threeparttable}
\usepackage{lineno}

\usepackage[T1]{fontenc}

\usepackage[utf8]{inputenc}

\usepackage{microtype}

\usepackage{inconsolata}

\usepackage{graphicx}
\usepackage{booktabs}
\usepackage{multirow}
\usepackage{multicol}
\usepackage{makecell}
\usepackage{ulem}

\title{\textsl{DR.EHR}: Dense Retrieval for Electronic Health Record with Knowledge Injection and Synthetic Data}

\author{Zhengyun Zhao \and Huaiyuan Ying \and Yue Zhong \and Sheng Yu\\
  Tsinghua University \\
  Beijing, China}

\begin{document}
\maketitle
\begin{abstract}
Electronic Health Records (EHRs) are pivotal in clinical practices, yet their retrieval remains a challenge mainly due to semantic gap issues. Recent advancements in dense retrieval offer promising solutions but existing models, both general-domain and biomedical-domain, fall short due to insufficient medical knowledge or mismatched training corpora. This paper introduces \texttt{DR.EHR}, a series of dense retrieval models specifically tailored for EHR retrieval. We propose a two-stage training pipeline utilizing MIMIC-IV discharge summaries to address the need for extensive medical knowledge and large-scale training data. The first stage involves medical entity extraction and knowledge injection from a biomedical knowledge graph, while the second stage employs large language models to generate diverse training data. We train two variants of \texttt{DR.EHR}, with 110M and 7B parameters, respectively. Evaluated on the CliniQ benchmark, our models significantly outperforms all existing dense retrievers, achieving state-of-the-art results. Detailed analyses confirm our models' superiority across various match and query types, particularly in challenging semantic matches like implication and abbreviation. Ablation studies validate the effectiveness of each pipeline component, and supplementary experiments on EHR QA datasets demonstrate the models' generalizability on natural language questions, including complex ones with multiple entities. This work significantly advances EHR retrieval, offering a robust solution for clinical applications.
\end{abstract}

\section{Introduction}
Electronic Health Records (EHRs) hold significant value in various clinical practices, and EHR retrieval plays a crucial role in enabling physicians to utilize EHRs more efficiently \citep{zhang2019high, ying2025geniegenerativenoteinformation}.
This step is essential in a wide range of clinical tasks, including patient cohort selection \citep{Jin2021AlibabaDA, Yang2021ImprovingCE}, EHR Question Answering (QA) \citep{Pampari2018emrQAAL, Lanz2024ParagraphRF}, and patient chart review \citep{gupta2024oncoretrievergenerativeclassifierretrieval, ye2021leveraging}.

Despite the critical importance of this field, its development has not progressed at a commensurate pace.
Most existing EHR retrieval systems, whether in academic research or deployed in real-world hospitals, still rely on exact match methods \citep{ruppel2020assessment, negro2021technological}, which inevitably suffer from the semantic gap issue \citep{koopman2016information, edinger2012barriers}.
A recent EHR retrieval benchmark, CliniQ \citep{cliniq}, which separately evaluates various matching types, quantitatively demonstrates that exact match methods struggle with semantic matches, even when augmented by query expansion using a Knowledge Graph (KG).

Recently, Dense Retrieval (DR), which leverages Pre-trained Language Models (PLMs) to generate dense text representations for retrieval, has garnered increasing research interest \citep{karpukhin2020dense}.
Owing to its inherent ability to capture semantics and large-scale contrastive learning, DR models have the potential to bridge the semantic gap and have exhibited strong zero-shot capabilities \citep{neelakantan2022text, bge_embedding}.
In the context of EHR retrieval, general-domain models such as \texttt{bge} \citep{bge_embedding} and \texttt{NV-Embed} \citep{lee2024nv} serve as strong baselines \citep{Myers2024LessonsLO}, but they leave significant room for improvement due to insufficient medical knowledge \citep{cliniq}.
Biomedical-domain models, including \texttt{MedCPT} \citep{jin2023medcpt} and \texttt{BMRetriever} \citep{xu2024bmretriever}, perform suboptimally despite ample knowledge, likely due to the mismatch between their training corpora and clinical notes.
Thus, there is a pressing need for an EHR dense retriever specifically designed for the task with comprehensive medical knowledge.

However, the development of an EHR retriever has been severely limited by the lack of training data \citep{jin2023medcpt, pmcp}.
The required query-document relevant pairs were traditionally accessible only through manual annotation.
The prohibitive costs of such annotations inevitably constrain the dataset scale to only dozens of queries, and the resulting models perform barely on par with BM25 \citep{Soni2020PatientCR}.
There have been attempts to generate large-scale relevance judgments automatically using string match algorithms or Large Language Models (LLMs) \citep{Shi2022ImprovingNM, gupta2024oncoretrievergenerativeclassifierretrieval}.
The increase in dataset scale leads to significant improvements in model performance.
Yet, the queries used in these works are still provided by human experts or fixed vocabularies, limiting the scale and diversity of the training data.
Consequently, the models lack generalizability and are only effective for specific diseases or even particular queries.

\begin{figure*}[tbp]
  \centering
  \includegraphics[width=0.95\textwidth]{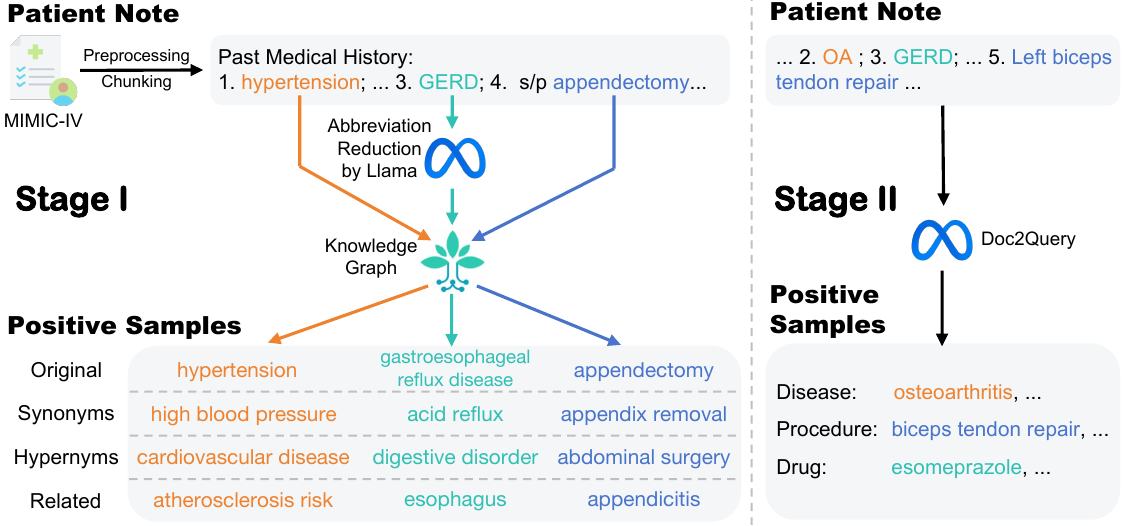}
  \caption{The training data collection pipeline of the two stages.
  In the first stage (left), the positive samples are defined as string-matched entities, reduced abbreviations, and their synonyms, hypernyms, and related entities sourced from the KG.
  In the second stage (right), the positive samples are generated by an LLM using Doc2Query.
  Note: OA is an abbreviation for osteoarthritis, and esomeprazole is generated since it is commonly used to treat GERD.}
  \label{fig:main}
\end{figure*}

In this work, we aim to develop a series of \textbf{\uline{D}}ense \textbf{\uline{R}}etrieval models for \textbf{\uline{E}}lectronic \textbf{\uline{H}}ealth \textbf{\uline{R}}ecord, dubbed \textbf{DR.EHR}.
Specifically, to address the need for extensive medical knowledge and generalizable models, we propose a two-stage training pipeline based on MIMIC-IV discharge summaries \citep{johnson2023mimic}.
In the first stage, we extract medical entity mentions from the EHRs and perform massive knowledge injection using a biomedical KG.
In the second stage, inspired by Doc2Query \citep{Nogueira2019DocumentEB}, we utilize LLMs to generate relevant entities for each EHR to collect large-scale and diverse training data.
The training data collection pipeline is summarized in Figure \ref{fig:main}.

We train two variants of \texttt{DR.EHR}, with 110M and 7B parameters, respectively, using contrastive learning with in-batch negatives.
On CliniQ, \texttt{DR.EHR-small} significantly outperforms all existing dense retrievers including 7B models, while our 7B variant demonstrates further improvement, achieving state-of-the-art results on the benchmark.
Detailed analysis demonstrates that the superiority of \texttt{DR.EHR} is substantial and consistent across different match types and query types.
Specifically, it achieves near-perfect performance on string matching and exhibits notable improvements on the most challenging semantic matching, such as implication and abbreviation matching.
Through extensive ablation studies, we validate the effectiveness of each component in the training pipeline, further substantiating the model's enhanced medical knowledge.
Though our models are trained exclusively on single-entity queries, they demonstrate strong generalizability over natural language questions including complex ones containing multiple entities.

Our contributions can be summarized as follows:
\begin{itemize}
    \item We propose a two-stage training pipeline using knowledge injection and synthetic data, addressing the lack of diverse training data of large scale and ample medical knowledge.
    \item We develop and release \texttt{DR.EHR}, a series of state-of-the-art and generalizable dense retrieval models specifically designed for the task of EHR retrieval.
    \item Detailed analysis demonstrate that \texttt{DR.EHR} overcomes the limitations of existing dense retrievers, exhibiting significantly richer medical knowledge and enhanced semantic matching capabilities.
\end{itemize}

\section{Related Work}

\subsection{EHR retrieval}
Most EHR retrieval methods rely on exact matches and heavily leverage biomedical KGs \citep{Hanauer2015SupportingIR,ruppel2020assessment}.
One popular approach to utilizing KGs for EHR retrieval is to identify medical entities in the EHRs and then match these entities with user queries \citep{Bonacin2018ExploringIO, Goodwin2017KnowledgeRA}. 
Other systems use KGs for query expansion.
By incorporating synonyms, abbreviations, and related concepts of user queries, these methods can significantly improve the recall rate \citep{Zhu2013UsingDS,alonso2016evaluation}. 
However, these methods are limited to exact matching and fixed vocabularies, and therefore struggle to process complex EHRs.

Constrained by the shortage of training data, only a limited number of studies have explored the application of supervised learning and language models in this field. 
\citet{Shi2022ImprovingNM} employed string matching to annotate the training data on imaging reports, and trained a dense retriever based on SentenceBERT \citep{Reimers2019SentenceBERTSE}. 
Despite its superiority on the leave-out test set, only hundreds of queries were incorporated, all focused on searching for diseases and anatomical findings in imaging reports.
Recently, \citet{gupta2024oncoretrievergenerativeclassifierretrieval} trained the Onco-Retriever series using a private dataset and annotations based on GPT-3.5, with model parameter sizes of 500M and 2B. 
On the manually annotated test set, Onco-Retriever outperformed the properitary model developed by OpenAI and \texttt{SFR-Embedding-Mistral} \citep{SFRAIResearch2024}. 
Yet, they only used 13 queries related to oncology, severely limiting the model's range of application.
Clearly, there is a lack of an EHR retriever that can effectively address the semantic match challenge and be applied to a wide range of queries.

\subsection{Knowledge injection}
Knowledge injection has been widely adopted as an effective approach to enriching the models' knowledge in the biomedical domain, primarily through KGs \citep{Trajanov2022ReviewON}.
Knowledge injection can be performed either during the pre-training phase or during fine-tuning for downstream tasks.
\citet{michalopoulos2020umlsbert} utilized UMLS, the most widely used biomedical KG, and introduced UmlsBERT.
By enhancing the model with semantic types of the entities and an additional prediction task for related entities, UmlsBERT demonstrated improvements across a variety of clinical tasks.
Others focus on obtaining better entity representations via knowledge injection and language models \citep{Yuan2020CODERKC, ying2024cortex}.
CODER \citep{Yuan2020CODERKC} employed contrastive learning on terms and relation triplets from UMLS to improve term normalization, significantly outperforming existing medical embeddings.
Similarly, \citet{Liu2020SelfAlignmentPF} introduced SapBERT, which used metric learning to cluster synonyms and achieved state-of-the-art results in medical entity linking tasks.

Knowledge injection has also been applied to dense retrieval.
\citet{Tan2023IncorporatingEK} fed an additional entity embedding sequence into the BERT model and used an entity similarity loss to inject knowledge into the model.
The resulting model, ELK, outperformed general domain retrievers in zero-shot biomedical retrieval tasks by a large margin.

\subsection{Synthetic data for retrieval}
Synthesizing data for retrieval may be traced back to Doc2Query \citep{Nogueira2019DocumentEB}, which was further expanded by \citet{Cheriton2019FromDT}.
The idea behind these methods was to generate pseudo queries for documents as document expansion.
With the rapid development of dense retrieval, training data soon became a scare resource, and research on synthetic data for retrieval turned to generate relevant queries from documents for model training.
\citet{Dai2022PromptagatorFD} utilized the FLAN model \citep{Wei2021FinetunedLM} to generate pseudo queries for each of the BEIR \citep{thakur2021beir} datasets.
\citet{Wang2023ImprovingTE} leveraged proprietary LLMs to generate diverse synthetic data across hundreds of thousands of tasks and 93 languages.
In the biomedical domain, \citet{xu2024bmretriever} also relied on proprietary LLMs and generated synthetic data for biomedicine.
So far, there has been no attempt to apply synthetic data for EHR retrieval.

\section{Methods}
We use MIMIC-IV discharge summaries as our training corpus.
Following \citet{cliniq}, we first clean the notes by removing all masks and excessive punctuation, and by converting all text to lowercase.
Then, we split all patient records into 100-word chunks with overlap of 10 words.
Based on this training corpus, we propose a two-stage training pipeline with synthetic data specifically designed for EHR retrieval.
The overall training data collection pipeline along with an example is demonstrated in Figure \ref{fig:main}.

\subsection{Stage I: Knowledge injection pre-training}
In the first stage, we aim to enrich the model's medical knowledge through contrastive learning.
For each note chunk used as an anchor, we first identify all entity mentions from it that are indexed in BIOS \citep{Yu2022BIOSAA}, the largest biomedical KG to date\footnote{We also tried UMLS, which yielded suboptimal results.}, as the initial positive sample set.
We only consider entities of specific semantic types to minimize noise included in the training data.
The list of incorporated semantic types can be found in Appendix \ref{app:type}.
Then, to further enhance the model's abilities to identify abbreviations, we prompt \texttt{Llama-3.1-8B-Instruct} to perform abbreviation reduction, and include the full names of the abbreviations appearing in the note as additional positive samples.
We conduct several cleaning steps to remove any noise generated by LLM and to ensure that the cleaned full names appear in BIOS.
The prompt used for abbreviation reduction and the detailed cleaning process are described in Appendix \ref{sec:abbr}.

Finally, as the core step to inject knowledge from the KG, we look up each positive entity in BIOS and incorporate their synonyms, hypernyms ("is a" relationship), and related entities (other relationships such as "may treat" and "may cause", full list in Appendix \ref{app:type}) into the positive sample set.
We do not include hyponyms ("reverse is a" relationship) since the information contained in the note is insufficient to deduce the hyponyms, and they will not be considered relevant in the downstream retrieval task.

In summary, given an anchor note chunk, its positive sample set consist of string-matched entities, full names of reduced abbreviations, and additional terms incorporated through BIOS.

\subsection{Stage II: Synthetic data fine-tuning}
In the second stage, we aim to fine-tune the model to optimize for the downstream EHR retrieval task using synthetic data.
Following CliniQ, we also focus on the task of entity retrieval, and consider three types of query entities: diseases, clinical procedures, and drugs.
We use \texttt{Llama-3.1-8B-Instruct} to generate various types of entities separately and combine them as the positive samples.
For better semantic matching capabilities, we prompt the LLM to generate entities that are either explicitly mentioned in or can be implicitly inferred from each note chunk.
The prompts used are provided in Appendix \ref{sec:syn}.

To understand the quality of LLM synthetic data, we conduct a manual evaluation of 50 randomly sampled note chunks containing 831 LLM-generated entities. 
This analysis revealed that 709 entities (85\%) are clinically validated by an M.D. candidate, demonstrating reasonable reliability of the data.
Most errors we identify are irrelevant medical entities with no explicit patterns.

\subsection{Model training}
We train two models of different sizes: \texttt{DR.EHR-small}, a BERT-based encoder with 110M parameters, initialized from \texttt{bge-base-en-v1.5}; and \texttt{DR.EHR-large}, a 7B decoder using the Mistral architecture, initialized from \texttt{NV-Embed-v2}.
These initialization choices are due to the superior performance of these models within their respective parameter sizes.
No middle-sized models are included since they generally perform worse than \texttt{bge-base}.

With different model architectures, the two models use distinct pooling strategies. 
For \texttt{DR.EHR-small}, we take the \texttt{[CLS]} embedding from the last layer as the text representation.
For \texttt{DR.EHR-large}, we adopt last token pooling.
The similarity $S(i,j)$ for an anchor $i$ and a sample $j$ is calculated as the cosine similarity of the two text embeddings.

In both stages, we train the model using Multi-Similarity Loss (MSL, \citealp{wang2019multi}) with in-batch negatives.
Formally, given an anchor $i$, its positive samples $\mathcal{P}(i)$, and its negative samples $\mathcal{N}(i)$, MSL first defines informative samples as follows:
\begin{equation}
    \mathcal{P}'(i) = \{j|j\in\mathcal{P}(i), S(i,j) < \max_{k\in\mathcal{N}(i)} S(i,k) + \epsilon\}
\end{equation}
\begin{equation}
    \mathcal{N}'(i) = \{j|j\in\mathcal{N}(i), S(i,j) > \min_{k\in\mathcal{P}(i)} S(i,k) - \epsilon\}
\end{equation}
where $\epsilon$ is a hyperparameter.
The loss for each anchor is calculated as follows:
\begin{equation}
    \begin{aligned}
        \mathcal{L} = & \frac{\log(1 + \sum_{j\in\mathcal{P}'(i)} \exp(-\alpha(S(i,j) - \lambda)))}{\alpha} \\
        & + \frac{\log(1 + \sum_{j\in\mathcal{N}'(i)} \exp(\beta(S(i,j) - \lambda)))}{\beta}
    \end{aligned}
\end{equation}
where $\alpha$, $\beta$, and $\lambda$ are hyperparameters.
In our experiments, we use $\epsilon=0.1, \alpha=2,\beta=50$, and $\lambda=0.5$, determined by grid search.

\section{Experiments}

\subsection{Statistics of the training data}
From the 332k discharge summaries in MIMIC-IV, we obtain over 5.8M note chunks for training, with an average of 17.5 chunks per note.
In the first training stage, the positive samples for each note chunk comprise three parts, with entities added from the KG further divided into three types: synonyms, hypernyms, and related entities.
For training efficiency, we only include at most two synonyms, two hypernyms, and two related entities for each positive entity sourced from string matching or abbreviation reduction.
For each hypernym or related entity included, we also additionally incorporate a random synonym of it, if any.
Consequently, for each positive entity, we add up to 10 terms from the KG.
In our pilot study, adding more entities did not lead to significant improvement.
Detailed statistics of these positive samples are presented in Table \ref{tab:stats}.
On average, each note chunk is associated with 137.9 positive samples, resulting in a total of over 802M samples.
Hypernyms, with an average of 50.9 samples per chunk, contribute the most, followed by related entities (38.6) and synonyms (30.2). 
Abbreviations account for the smallest proportion, with only 2.4 per chunk, and nearly 28\% of chunks contain no abbreviations.

\begin{table}[htbp]
\centering
  \caption{Statistics of positive samples for each chunk used in the first training stage.
  Avg: average; Q1: first quartile; Q3: third quartile; KG: knowledge graph.}
  \label{tab:stats}
  \resizebox{\columnwidth}{!}{
  \begin{tabular}{lccccc}
    \toprule
    \textbf{Source} & \textbf{Avg} & \textbf{Q1} & \textbf{Q3} & \textbf{Max} & \textbf{Sum}\\
    \midrule
    String Match & 15.7 & 12 & 20 & 64 & 91M\\
    Abbreviation & 2.4 & 0 & 3 & 25 & 14M\\
    KG\\
    \quad Synonym & 30.2 & 22 & 38 & 127 & 176M\\
    \quad Hypernym & 50.9 & 38 & 64 & 185 & 296M\\
    \quad Related & 38.6 & 25 & 51 & 216 & 225M\\
    \midrule
    Overall & 137.9 & 102 & 172 & 588 & 802M\\
    \bottomrule
  \end{tabular}}
\end{table}

In the second training stage, the number of positive samples generated is significantly less than that in the first stage.
Detailed statistics, categorized by entity type, are provided in Table \ref{tab:stats2}.
On average, each chunk has 15.8 positive samples generated by the LLM, resulting in a total of nearly 86M samples. 
The generated entities exhibit a relatively even distribution among the three entity types.

\begin{table}[tbp]
\centering
  \caption{Statistics of positive samples for each chunk used in in the second training stage.
  Avg: average; Q1: first quartile; Q3: third quartile.}
  \label{tab:stats2}
  \begin{threeparttable}
  \begin{tabular}{lccccc}
    \toprule
    \textbf{Entity Type} & \textbf{Avg} & \textbf{Q1} & \textbf{Q3} & \textbf{Max} & \textbf{Sum}\\
    \midrule
    Disease & 5.4 & 3 & 7 & 33 & 26M\\
    Procedure & 7.3 & 5 & 9 & 31 & 42M\\
    Drug & 4.6 & 2 & 6 & 32 & 20M\\
    \midrule
    Overall & 15.8\tnote{*} & 11 & 20 & 63 & 86M\\
    \bottomrule
  \end{tabular}
  \begin{tablenotes}  
    \footnotesize  
    \item[*] The LLM may generate repeated entities in three rounds so the combined count is less than the sum of three types.
  \end{tablenotes}  
  \end{threeparttable}
\end{table}

\subsection{Model training}
In our experiments, the maximum token length is set to 512 for note chunks and 16 for entities.
To facilitate batch training, we up-sample or down-sample the positive entities of each chunk to a fixed number.
We employ distinct data allocation strategies for the two models across two training stages, due to the different GPU memory requirements of the models and the varying dataset scales for each stage.
\texttt{DR.EHR-large} is trained with less data due to the higher GPU memory constraints.
The hyperparameters and details in the training process are presented in Appendix \ref{app:train}.

\begin{table*}[htbp]
  \centering
  \caption{Performance of various dense retrievers on CliniQ.
  QE: Query expansion.
  Dim: Dimension of the embeddings.
  R@100: Recall at 100.
  The \textbf{bold} and \underline{underlined} values represent the best and second-best results, respectively, in each column.
  }
  \label{tab:overall}
  \resizebox{\textwidth}{!}{
  \begin{threeparttable}
  \begin{tabular}{lcccccccc}
    \toprule
    \multirow{2}{*}[-0.8ex]{\textbf{Model}} & \multirow{2}{*}[-0.8ex]{\textbf{Size}} & \multirow{2}{*}[-0.8ex]{\textbf{Dim}} & \multicolumn{3}{c}{\textbf{Single-Patient}} & \multicolumn{3}{c}{\textbf{Multi-Patient}}\\
    \cmidrule(r){4-6} \cmidrule(r){7-9}
    & & & \textbf{MRR} & \textbf{NDCG} & \textbf{MAP} & \textbf{MRR} & \textbf{NDCG@10} & \textbf{R@100}\\
    \midrule
    \texttt{bge-base-en-v1.5} & 110M & 768 & 82.48 & 83.59 & 74.54 & 54.97 & 56.51 & 39.50\\
    \texttt{MedCPT} & 220M\tnote{*} & 768 & 84.23 & 85.49 & 77.42 & 47.21 & 50.07 & 41.97\\
    \texttt{text-embedding-3-large} & - & 3072 & 85.16 & 86.09 & 78.36 & 59.54 & 60.45 & 48.75\\
    \texttt{gte-Qwen2-7B-Instruct} & 7B & 3584 & 84.59 & 85.33 & 77.02 & 60.39 & 62.06 & 48.04\\
    \texttt{NV-Embed-v2} & 7B & 4096 & 86.57 & 87.36 & 80.21 & 59.48 & 62.06 & 51.54\\
    \midrule
    \texttt{DR.EHR-small} & 110M & 768 & \underline{92.96} & \textbf{93.26} & \textbf{89.12} & \underline{67.06} & \underline{68.75} & \underline{64.11}\\
    \quad w/o stage I & 110M & 768 & 91.61 & 92.00 & 87.15 & 65.55 & 67.59 & 60.42\\
    \texttt{DR.EHR-large}& 7B & 4096 & \textbf{93.03} & \underline{93.20} & \underline{88.94} & \textbf{68.97} & \textbf{71.34} & \textbf{67.04}\\
    \bottomrule
  \end{tabular}
  \begin{tablenotes}  
    \footnotesize  
    \item[*] \texttt{MedCPT} has separate query encoder and document encoder, so we count the parameter size as the summation of both models.
  \end{tablenotes}  
  \end{threeparttable}}
\end{table*}

\subsection{Model evaluation}
We mainly evaluate our models on CliniQ, the only publicly available EHR retrieval benchmark of large scale.
CliniQ is constructed with 1k patient summaries from MIMIC-III, split into 16.5k chunks of 100 words each.
It contains over 1k queries of three types: diseases, clinical procedures, and drugs, collected from structured codes in MIMIC and annotated by GPT-4o.
It incorporates two retrieval settings: Single-Patient retrieval where models are tasked with ranking the chunks of a single patient note given a query, and Multi-Patient retrieval, where model are required to retrieve relevant chunks from the entire set of 16.5k chunks.
On Single-Patient Retrieval, models are evaluated with Mean Reciprocal Rank (MRR), Normalized Discounted Cumulative Gain (NDCG), and Mean Average Precision (MAP).
On Multi-Patient Retrieval, models are evaluated with MRR, NDCG at 10, and recall at 100.
CliniQ provides additional semantic match assessment by further classifying the relevance judgments into various categories under the Single-Patient Retrieval setting.
With this comprehensive benchmark, we are able to assess the performance of \texttt{DR.EHR} under various clinical scenarios.

In addition to CliniQ, which comprises only entity-based queries, we evaluate our models by adapting existing EHR QA datasets into a retrieval framework to demonstrate our models' generalizability on natural language queries.
However, since these datasets are not rigorous retrieval benchmarks that have undergone peer reviews, we only present them as a supplement to our main experiments in Appendix \ref{sec:qa}.
To briefly summarize, we employ the Single-Patient Retrieval setting and split the notes into chunks. 
For each QA dataset question, the models are tasked with ranking the chunk containing the correct answer highest among all chunks from that patient's note, assessed using MRR.

\subsection{Main results}
The performance of \texttt{DR.EHR} on CliniQ is presented in Table \ref{tab:overall}, in comparison with \texttt{bge-base-en-v1.5}, \texttt{MedCPT}, \texttt{text-embedding-3-large} by OpenAI, \texttt{gte-Qwen2-7B-Instruct} \citep{Li2023TowardsGT}, and \texttt{NV-Embed-v2}.
Our proposed models present superior performance on CliniQ.
Specifically, \texttt{DR.EHR-small} with 110M parameters outperforms all existing dense retrievers, including the proprietary embedding model by OpenAI and state-of-the-art 7B models, by a remarkable margin.
The large variant with 7B parameters achieves further significant improvement on Multi-Patient Retrieval.
The advantages of \texttt{DR.EHR} are consistent and substantial across both retrieval settings and all metrics.
Notably, we improve the MAP on Single-Patient Retrieval from the previous SOTA of 80.21 to 89.12 for \texttt{DR.EHR-small} and 88.94 for \texttt{DR.EHR-large}, and the Recall@100 on Multi-Patient Retrieval from the previous SOTA of 51.54 to 64.11 for \texttt{DR.EHR-small} and 67.04 for \texttt{DR.EHR-large}.

The performance difference between the two variants of the \texttt{DR.EHR} models is more pronounced in the Multi-Patient Retrieval setting, likely due to its inherently greater complexity. 
With its larger parameter size and higher-dimensional embeddings, \texttt{DR.EHR-large} captures more nuanced medical knowledge and produces more discriminative representations. 
However, this advantage may be less noticeable in the Single-Patient setting, where the task is comparatively simpler with only 16.6 chunks per query to be ranked.

\begin{table*}[htbp]
\centering
  \caption{Performance of various dense retrievers and ablation study on Single-Patient Retrieval, dissected by match types.
  The score for each type is the average of MRR, NDCG, and MAP.
  In the ablation study part, "w/o stage I" indicates the removal of stage I training, and each row starting with "+" represents adding extra training data in stage I to the previous row, with the same  training data split as in Table \ref{tab:stats}.
  The \textbf{bold} and \underline{underlined} values represent the best and second-best results, respectively, in each column.
  }
  \label{tab:match}
  \begin{tabular}{lccccc}
    \toprule
    \textbf{Model} & \textbf{String} & \textbf{Synonym} & \textbf{Abbreviation} & \textbf{Hyponym} & \textbf{Implication}\\
    \midrule
    \texttt{bge-base-en-v1.5} & 86.75 & 71.57 & 57.15 & 64.42 & 52.75\\
    \texttt{NV-Embed-v2} & 87.34 & 83.28 & 72.13 & 75.07 & 59.96\\
    \midrule
    \texttt{DR.EHR-small} & 97.34 & \underline{86.01} & \underline{83.37} & \textbf{76.88} & \textbf{67.56}\\
    \quad w/o stage I & 97.27 & 82.13 & 78.31 & 71.06 & 63.91\\
    \quad + String Match & 97.60 & 81.37 & 78.26 & 70.23 & 63.18\\
    \quad + Abbreviation & 97.47 & 81.69 & 80.40 & 69.98 & 63.96\\
    \quad + KG--Synonym & \textbf{97.66} & 84.07 & 80.79 & 71.31 & 64.35\\
    \quad + KG--Hypernym & 97.42 & 85.87 & 81.86 & \underline{75.71} & 64.19\\
    \texttt{DR.EHR-large} & \underline{97.61} & \textbf{86.27} & \textbf{85.08} & 74.99 & \underline{65.35}\\
    \bottomrule
  \end{tabular}
\end{table*}

\section{Analysis}

\subsection{Semantic match assessment}
\label{sec:match}
The performance of various models on semantic match assessment in CliniQ are presented in Table \ref{tab:match}.
For brevity, we only report the average score of MRR, NDCG, and MAP.
Full results can be found in Appendix \ref{app:detail}.
\texttt{DR.EHR} demonstrates significant improvements over the baseline models.
Specifically, \texttt{DR.EHR} addresses the challenge of insufficient exact match capabilities observed in general-domain dense retrievers \citep{Zhuang2023TyposawareBP} in the context of EHR retrieval, achieving near-perfect performance on the string match benchmark in CliniQ, compared to around 87\% of the baselines.

In terms of semantic matches, \texttt{DR.EHR-small} outperforms its initialization model by more than 10\% across all categories, with a notable improvement of over 26\% in abbreviation matching.
These substantial gains underscore the effectiveness of the proposed pipeline.
Through extensive knowledge injection and synthesized data tailored for this task, the models have learned to capture deep semantic associations between terms and to represent them effectively in their embeddings.
The two proposed models of distinct parameter sizes do not show noticeable difference under the Single-Patient Retrieval setting.

\subsection{Query type assessment}
The model performances for different query types (disease, procedure, and drug) are presented in Table \ref{tab:query}.
As in the previous section, we only report average scores and leave the full results in Appendix \ref{app:detail}.
We additionally include the BM25 baseline, which achieves the best performance for drug searches in Multi-Patient Retrieval.
The superiority of BM25 on this benchmark may be attributed to the fact that most drug queries consist of single words that appear verbatim in the notes.
\texttt{DR.EHR} demonstrates consistent and significant improvements across all query types.
Notably, it addresses the limitations of other dense retrievers in drug matching, improving the average scores by 12\% and 24\% in the two retrieval settings, respectively.
This improvement can be related to the enhanced string match abilities in Section \ref{sec:match}.
Consistent with previous findings, \texttt{DR.EHR-large} presents its advantage only in Multi-Patient Retrieval.

\begin{table*}[htbp]
\centering
  \caption{Performance of various retrieval methods and ablation study for different query types.
  The score for each type is the average of MRR, NDCG, and MAP in Single-Patient Retrieval, and the average of MRR, NDCG@10, and Recall@100 in Multi-Patient Retrieval.
  In the ablation study part, "Stage I +" indicates using only the specific type of synthesized data for training during stage II.
  The \textbf{bold} and \underline{underlined} values represent the best and second-best results, respectively, in each column.}
  \label{tab:query}
  \begin{tabular}{lcccccc}
    \toprule
    \multirow{2}{*}[-0.7ex]{\textbf{Model}} & \multicolumn{3}{c}{\textbf{Single-Patient}} & \multicolumn{3}{c}{\textbf{Multi-Patient}}\\
    \cmidrule(r){2-4} \cmidrule(r){5-7}
    & \textbf{Disease} & \textbf{Procedure} & \textbf{Drug} & \textbf{Disease} & \textbf{Procedure} & \textbf{Drug}\\
    \midrule
    BM25 & 64.69 & 64.81 & 72.08 & 33.76 & 33.55 & 76.91\\
    \texttt{bge-base-en-v1.5} & 75.98 & 75.83 & 82.47 & 40.46 & 41.48 & 62.06\\
    \texttt{NV-Embed-v2} & 81.95 & 82.82 & 86.04 & 51.50 & 54.11 & 63.76\\
    \midrule
    \texttt{DR.EHR-small} & \textbf{87.52} & \underline{83.95} & \underline{94.61} & \underline{51.58} & \underline{50.90} & \underline{86.03}\\
    \quad Stage I + Disease & 83.74 & 79.34 & 82.99 & 49.60 & 44.49 & 60.15\\
    \quad Stage I + Procedure & 83.89 & 82.03 & 92.19 & 45.15 & 48.86 & 80.75\\
    \quad Stage I + Drug & 73.18 & 71.61 & 91.29 & 33.28 & 32.14 & 85.49\\
    \texttt{DR.EHR-large} & \underline{86.50} & \textbf{84.99} & \textbf{94.71} & \textbf{54.37} & \textbf{52.65} & \textbf{88.89}\\
    \bottomrule
  \end{tabular}
\end{table*}

\subsection{Ablation study}
We conduct three ablation studies using \texttt{DR.EHR-small}.
First, we ablate the stage I training and present the results in Tables \ref{tab:overall} and \ref{tab:match}.
The results demonstrate that the knowledge injection phase significantly contributes to the final performance of \texttt{DR.EHR}, particularly on Recall@100 for Multi-Patient Retrieval.
Detailed analysis of different match types reveals that this contribution is primarily attributed to semantic matches.
The knowledge injection phase improves model performance by approxiamately 5\% across all semantic match types.

To gain a deeper understanding of the contributions of knowledge injection, we divide the Stage I training data into five parts, as shown in Table \ref{tab:stats}, and sequentially incorporate each part to demonstrate their individual effects.
The results, presented in Table \ref{tab:match}, demonstrate that each portion of the training data significantly enhances performance on the corresponding benchmark, confirming that \texttt{DR.EHR} effectively acquires extensive knowledge from KGs.
Notably, the additional training data also improves performance on other types of matching in most cases, indicating enhanced generalizability of \texttt{DR.EHR}.

For the second stage training, we divide the synthetic data according to the generated query types, and use them separately to train a series of models.
Results in Table \ref{tab:query} demonstrate that synthetic data of specific query type improves model performance on the corresponding benchmark.
Surprisingly, however, combining various types of synthetic data further enhances model capabilities significantly across all query types compared to models trained on individual data types.
This synergistic effect of "1+1+1>3" might suggest that our models benefit from transfer learning during the second stage of training. 
When exposed to diverse query types, \texttt{DR.EHR} learns to capture broader semantic patterns and deeper knowledge connections, resulting in enhanced generalization capabilities and improved learning efficiency.

\subsection{Case study}
We conduct several case studies comparing \texttt{bge-base-en-v1.5} and \texttt{DR.EHR-small}.
For each match type, one example is selected, and the queries, note chunks, corresponding ranks, and cosine similarities generated by the two models are provided in Appendix \ref{sec:case}.
The rank is calculated after excluding relevant chunks of other match types, and the cosine similarity is computed between the query and the relevant part (see Table \ref{tab:case}) within the chunks.
Our observations reveal that \texttt{DR.EHR-small} successfully identifies various types of matches, and its higher cosine similarities demonstrate its ability to learn extensive medical knowledge and represent information in clinical notes more effectively.

\subsection{Generalizability assessment}
The training pipeline and prior evaluations on CliniQ focus exclusively on single-entity queries.
To assess the generalizability of our models, we conduct additional experiments on adapted EHR QA datasets featuring natural language questions, including a dedicated subset comprising exclusively complex, multi-entity queries.
The results in Appendix \ref{sec:qa} reveal that \texttt{DR.EHR} maintains substantial advantages over baseline models across most evaluated datasets.
This indicates strong generalization capability to unseen query types, even when trained exclusively on entity-based queries.

\section{Conclusion}
In this paper, we propose a two-stage training pipeline specifically designed for the task of EHR retrieval.
The first stage employs KGs for knowledge injection while the second stage fine-tunes the model for the retrieval task with LLM synthetic data.
Using this pipeline, we develop and release \texttt{DR.EHR}, a state-of-the-art EHR retriever available in two model sizes.
Extensive experiments demonstrate that \texttt{DR.EHR} significantly outperforms baseline models across various settings, match types, and query types.
Notably, our models overcome challenges faced by traditional dense retrievers and exhibit exceptional capabilities in both string matching and semantic matching.
Supplementary experiments further verify the models' generalizability to complex natural language queries.

\section*{Limitations}
This study has several limitations.
First, the evaluation of our model mainly focus on CliniQ, specifically the task of entity retrieval.
Due to the lack of other public EHR retrieval benchmarks, we utilize EHR QA datasets to illustrate the generalizability of our models.
However, these datasets are not rigorous retrieval benchmarks, and we call for future efforts to construct richer and more diverse public benchmarks.
Second, the quality of LLM synthetic data could be improved, as pointed out by our manual inspection.
Nevertheless, removing the noises involves differentiating between relevant and irrelevant medical entities and generally requires more advanced LLMs, which is computationally intensive and exceeds our current resource constraints given the enormous scale of our data.
Third, while hard negatives are known to significantly enhance model performance, particularly during task-specific fine-tuning \citep{karpukhin2020dense, Zeng2022AutomaticBT}, the design of synthetic hard negative data is non-trivial. We leave this challenge for future research.

\bibliography{custom}

\appendix

\section{Included semantic types and relationship types}
\label{app:type}
List of semantic types in BIOS included in the training data:
"Laboratory Procedure", "Sign, Symptom, or Finding", "Diagnostic Procedure", "Therapeutic or Preventive Procedure", "Disease, Syndrome or Pathologic Function", "Chemical or Drug".

List of relationships included in the training data:
"may be treated by", "may treat", "may be diagnosed by", "may diagnose", "may be caused by", "may cause".

\section{Details of Abbreviation Reduction}
\label{sec:abbr}
\begin{figure}[t]
  \includegraphics[width=\columnwidth]{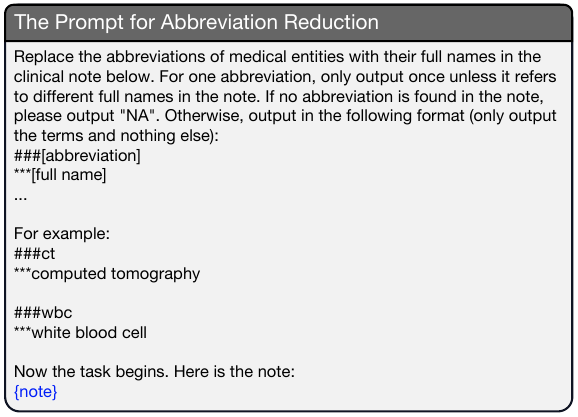}
  \caption{The prompt used for abbreviation reduction.
  \textcolor{blue}{\{note\}} is the placeholder for the note to be processed.}
  \label{fig:prompt_abbr}
\end{figure}

The prompt used for abbreviation reduction is provided in Figure \ref{fig:prompt_abbr}.
After reducing abbreviations, we conduct the following cleaning steps to eliminate potential noise generated by the LLM:
\begin{enumerate}
    \item We remove abbreviations that do not appear in the original note.
    \item We remove full names that are identical to their abbreviations.
    \item We remove full names that are not indexed in BIOS.
    \item We remove abbreviations that are only one character long.
\end{enumerate}

\section{Prompt for synthetic data generation}
\label{sec:syn}
\begin{figure}[t]
  \includegraphics[width=\columnwidth]{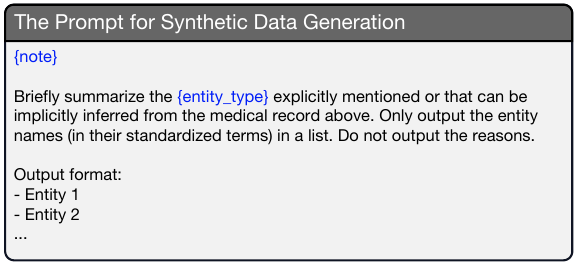}
  \caption{The prompt used for synthetic data generation.
  \textcolor{blue}{\{note\}} is the placeholder for the note to be processed, and \textcolor{blue}{\{entity\_type\}} takes on the values of diseases, clinical procedures, and drugs.}
  \label{fig:prompt_syn}
\end{figure}

The prompt used for synthetic data generation is given in Figure \ref{fig:prompt_syn}.

\section{Details in the training process}
\label{app:train}
The models are trained using 8 Nvidia A800 GPUs.
Following \citet{lee2024nv}, \texttt{DR.EHR-large} is trained using low-rank adaptation (LoRA, \citealp{hu2021lora}) with rank 16, alpha 32 and a dropout rate of 0.1.
To further reduce GPU memory requirements, techniques including Bfloat 16 training and DeepSpeed ZeRO-2 are applied to \texttt{DR.EHR-large}.
All training processes are optimized using AdamW \citep{Loshchilov2017DecoupledWD} with default parameters and a learning rate of 1e-4.
We set a warmup ratio of 0.1 and a linear decay for the learning rate scheduler.
The data-related hyperparameters for different models across the two stages are shown in Table \ref{tab:hyper}.

\begin{table}[htbp]
\centering
  \caption{Data-related hyperparameters used for different models across different training stages.
  Pos: the number of positive samples per chunk.}
  \label{tab:hyper}
  \begin{threeparttable}
  \begin{tabular}{ccccc}
    \toprule
    \textbf{Stage} & \textbf{Model} & \textbf{Pos} & \textbf{Batch Size}\tnote{*} & \textbf{Epoch}\\
    \midrule
    \multirow{2}{*}{I} & small & 128 & 32 & 3\\
    & large & 32 & 16 & 1\\
    \midrule
    \multirow{2}{*}{II} & small & 16 & 32 & 1\\
    & large & 16 & 16 & 1\\
    \bottomrule
  \end{tabular}
  \begin{tablenotes}  
    \footnotesize  
    \item[*] With in-batch negatives, the ratio of positive to negative samples is batch size minus one.
  \end{tablenotes}  
  \end{threeparttable}
\end{table}

\section{Additional evaluation on EHR QA datasets}
\label{sec:qa}
To illustrate our models' generalizability on different types of queries, we utilize existing EHR QA datasets, specifically, WhyQA \citep{fan-2019-annotating}, emrQA \citep{pampari-etal-2018-emrqa}, and RxWhyQA \citep{Moon_2023}, all derived from n2c2 challenges.\footnote{https://n2c2.dbmi.hms.harvard.edu/}
These datasets consist of natural language questions with answers extracted from clinical notes. 
For our evaluation, we adopt the Single-Patient Retrieval setting in CliniQ.
We process one note at a time, and segment the note into 100-word chunks.
Given each question, the models are required to rank the chunk containing the correct answer highest.

For each dataset, we select only questions with a single definitive answer and source notes exceeding 500 words. 
Where multiple rephrased versions of a question existed, we randomly sample one. 
Additionally, we use GPT-4o-mini to identify complex questions involving multiple entities. 
In total, we collect 19,121 questions, of which 2,630 (14\%) contained multiple entities. 
The detailed dataset statistics are presented in Table \ref{tab:qa_stat}.

\begin{table}[htbp]
\centering
  \caption{Statistics of the benchmarks adapted from various EHR QA datasets.
  Complex: complex questions with multiple queries identified by GPT-4o-mini.}
  \label{tab:qa_stat}
  \resizebox{\columnwidth}{!}{
  \begin{threeparttable}
  \begin{tabular}{lccc}
    \toprule
    \textbf{Dataset} & \textbf{Notes} & \textbf{Questions} & \textbf{Complex}\\
    \midrule
    WhyQA & 133 & 237 & 30\\
    emrQA\tnote{*}\\
    \quad risk & 116 & 1,209 & 53\\
    \quad relations & 324 & 6,750 & 1,879\\
    \quad medication & 257 & 7,495 & 153\\
    RxWhyQA & 283 & 3,430 & 515\\
    \midrule
    \textbf{Total} & 1,113 & 19,121 & 2,630\\
    \bottomrule
  \end{tabular}
  \begin{tablenotes}  
    \footnotesize  
    \item[*] emrQA has five subsets, while the other two (smoking and obesity) lack diverse questions.
  \end{tablenotes}  
  \end{threeparttable}
  }
\end{table}

We evaluate the Mean Reciprocal Rank (MRR) on these test sets (full set and multi-entity subset) for DR.EHR models compared to their backbone models, as shown in Table \ref{tab:qa_re} and \ref{tab:qa_mul}.
The results demonstrate that DR.EHR consistently outperforms its base models across most datasets, both in the full sets and the multi-entity subsets, confirming its generalizability to natural language questions and complex multi-entity queries.

\begin{table*}[htbp]
\centering
  \caption{Performance (MRR) of \texttt{DR.EHR} on the benchmarks adapted from EHR QA datasets, in comparison to their backbone models.
  The \textbf{bold} and \underline{underlined} values represent the best and second-best results, respectively, in each column.}
  \label{tab:qa_re}
  \resizebox{\linewidth}{!}{
  \begin{tabular}{lccccc}
    \toprule
    \textbf{Model} & \textbf{WhyQA} & \textbf{emrQA-risk} & \textbf{emrQA-relations} & \textbf{emrQA-medication} & \textbf{RxWhyQA}\\
    \midrule
    \texttt{bge-base-en-v1.5}	&79.85&	32.35	&70.01&	73.42&	66.14\\
    \texttt{NV-Embed-v2}&	84.83&	\textbf{46.58}	&72.42&	81.11&	77.28\\
    \texttt{DR.EHR-small}	&\underline{85.44} &43.84 &	\underline{77.46} &	\underline{82.23}&	\underline{84.70}\\
    \texttt{DR.EHR-large}&	\textbf{87.55} &	\underline{44.56} &	\textbf{78.24} &	\textbf{83.12}&\textbf{85.19} \\
    \bottomrule
  \end{tabular}
  }
\end{table*}

\begin{table*}[htbp]
\centering
  \caption{Performance (MRR) of \texttt{DR.EHR} on the multi-query subsets of the benchmarks adapted from EHR QA datasets, in comparison to their backbone models.
  The \textbf{bold} and \underline{underlined} values represent the best and second-best results, respectively, in each column.}
  \label{tab:qa_mul}
  \resizebox{\linewidth}{!}{
  \begin{tabular}{lccccc}
    \toprule
    \textbf{Model} & \textbf{WhyQA} & \textbf{emrQA-risk} & \textbf{emrQA-relations} & \textbf{emrQA-medication} & \textbf{RxWhyQA}\\
    \midrule
    \texttt{bge-base-en-v1.5}	&87.61	&48.89&	71.47	&75.82&	71.52\\
    \texttt{NV-Embed-v2}&	84.92	&\underline{60.12}	&72.43&	\textbf{79.03}	&78.53\\
    \texttt{DR.EHR-small}	&\underline{88.15} &	52.84 &	\underline{75.79}&75.92&	\underline{87.51} \\
    \texttt{DR.EHR-large}&	\textbf{90.83}&	\textbf{62.79} &	\textbf{78.57} &\underline{76.72}&	\textbf{89.78} \\
    \bottomrule
  \end{tabular}
  }
\end{table*}

However, these benchmarks suffer from several limitations.
First, these datasets are originally designed for QA systems, and their effectiveness in evaluating retrieval models remains unverified.
Second, most questions in these datasets focus on entities with exact string matches in the notes, as they were constructed via entity extraction and template filling. 
Consequently, they are barely satisfactory in assessing dense retrievers.
Third, the template-generated questions exhibit limited diversity in question formulation, limiting their representativeness.

\section{Full results of semantic match and query type assessment}
\label{app:detail}
Full results of the semantic match assessment is provided in Table \ref{tab:match_full}, and metrics for the query type assessment is provided in Table \ref{tab:query_full}.

\begin{table*}[htbp]
  \caption{Performance of various retrieval methods on Single-Patient Retrieval, dissected by match types.}
  \label{tab:match_full}
  \centering
  \begin{tabular}{lcccc}
    \toprule
    \textbf{Model} & \textbf{Match Type} & \textbf{MRR} & \textbf{NDCG} & \textbf{MAP}\\
    \midrule
    \texttt{bge-base-en-v1.5} & String & 87.35 & 88.93 & 83.96\\
    & Synonym & 72.48 & 76.45 & 65.78\\
    & Abbreviation & 55.15 & 64.55 & 51.74\\
    & Hyponym & 63.34 & 70.52 & 59.41\\
    & Implication & 51.70 & 61.05 & 45.51\\
    \midrule
    \texttt{NV-Embed-v2} & String & 87.67 & 89.50 & 84.85\\
    & Synonym & 84.29 & 86.17 & 79.37\\
    & Abbreviation & 71.50 & 76.91 & 67.97\\
    & Hyponym & 74.41 & 79.39 & 71.40\\
    & Implication & 59.59 & 67.04 & 53.25\\
    \midrule
    \texttt{DR.EHR-small} & String & 97.32 & 97.81 & 96.88\\
    & Synonym & 86.44 & 88.53 & 83.05\\
    & Abbreviation & 82.63 & 86.31 & 81.16\\
    & Hyponym & 75.76 & 81.02 & 73.85\\
    & Implication & 67.40 & 73.37 & 61.91\\
    \midrule
    \texttt{DR.EHR-large} & String & 97.62 & 98.03 & 97.19\\
    & Synonym & 86.80 & 88.72 & 83.28\\
    & Abbreviation & 84.51 & 87.73 & 82.99\\
    & Hyponym & 74.08 & 79.42 & 71.48\\
    & Implication & 65.15 & 71.54 & 59.37\\
    \bottomrule
  \end{tabular}
\end{table*}

\begin{table*}[htbp]
  \caption{Performance of various retrieval methods on different types of queries.}
  \label{tab:query_full}
  \centering
  \resizebox{\linewidth}{!}{
  \begin{tabular}{lccccccc}
    \toprule
    \multirow{2}{*}[-0.7ex]{\textbf{Model}} & \multirow{2}{*}[-0.6ex]{\textbf{\makecell[c]{Query\\Type}}} & \multicolumn{3}{c}{\textbf{Single-Patient}} & \multicolumn{3}{c}{\textbf{Multi-Patient}}\\
    \cmidrule(r){3-5} \cmidrule(r){6-8}
    & & \textbf{MRR} & \textbf{NDCG} & \textbf{MAP} & \textbf{MRR} & \textbf{NDCG@10} & \textbf{Recall@100}\\
    \midrule
    BM25 & Disease & 68.01 & 71.42 & 54.65 & 39.62 & 41.26 & 20.41\\
    & Procedure & 66.75 & 71.12 & 56.55 & 32.57 & 34.54 & 33.55\\
    & Drug & 73.78 & 76.22 & 66.24 & 88.43 & 88.59 & 53.70\\
    \midrule
    \texttt{bge-base-en-v1.5} & Disease & 80.05 & 80.49 & 67.41 & 44.23 & 47.35 & 29.80\\
    & Procedure & 78.71 & 80.12 & 68.66 & 38.15 & 41.49 & 44.81\\
    & Drug & 83.99 & 85.29 & 78.12 & 72.49 & 71.91 & 41.77\\
    \midrule
    \texttt{NV-Embed-v2} & Disease & 85.83 & 85.28 & 74.75 & 55.16 & 58.32 & 41.01\\
    & Procedure & 85.20 & 85.88 & 77.38 & 48.70 & 52.78 & 60.86\\
    & Drug & 87.06 & 88.37 & 82.69 & 69.19 & 70.42 & 51.66\\
    \midrule
    \texttt{DR.EHR-small} & Disease & 90.14 & 89.94 & 82.47 & 52.63 & 55.50 & 46.60\\
    & Procedure & 85.98 & 86.87 & 79.01 & 43.86 & 47.37 & 61.47\\
    & Drug & 95.12 & 95.51 & 93.20 & 91.00 & 90.78 & 76.30\\
    \midrule
    \texttt{DR.EHR-large} & Disease & 89.42 & 89.11 & 81.02 & 55.06 & 58.12 & 49.94\\
    & Procedure & 86.89 & 87.76 & 80.39 & 44.27 & 49.62 & 63.57\\
    & Drug & 95.35 & 95.59 & 93.24 & 93.60 & 93.59 & 79.52\\
    \bottomrule
  \end{tabular}
    }
\end{table*}

\section{Case studies}
\label{sec:case}
We present several cases in Table \ref{tab:case} where \texttt{bge-base-en-v1.5} fails to retrieve the relevant chunk, while \texttt{DR.EHR} succeeds.
One example is provided for each match type.

\begin{table*}[htbp]
\begin{threeparttable}
\centering
  \caption{Case studies of the performance of \texttt{DR.EHR} compared to \texttt{bge-base-en-v1.5} on Single-Patient Retrieval.
  The last two columns are the rank of the corresponding chunk and the cosine similarity given by the two models.
  The rank is calculated after removing relevant chunks of other match types.
  The cosine similarity is between the query and the relevant part (in red).}
  \label{tab:case}
  \begin{tabular}{lllcc}
    \toprule
    \textbf{Match Type} & \textbf{Query} & \textbf{Patient note} & \textbf{\texttt{bge}} & \textbf{\texttt{DR.EHR}}\\
    \midrule
    String & ceftriaxone & \makecell[l]{... She was given Vanc, \textbf{\textcolor{red}{Ceftriaxone}}, Flagyl,\\2L IVF, and started on levophed ...} & 12 / 1.00 & 1 / 1.00\\
    \midrule
    Synonym & phenytoin & \makecell[l]{... MEDICINE Allergies: \textbf{\textcolor{red}{Dilantin}}\tnote{1} ...} & 7 / 0.61 & 1 / 0.86\\
    \midrule
    Abbreviation & hypertension & \makecell[l]{...Past Medical History: (1) \textbf{\textcolor{red}{HTN}}\tnote{2} (2) ...} & 15 / 0.61 & 1 / 0.89\\
    \midrule
    Hyponym & \makecell[l]{interruption of\\the vena cava} & \makecell[l]{... Prophylaxis: \textbf{\textcolor{red}{IVC filter}}\tnote{3} and Pneumoboots. ...} & 5 / 0.59 & 1 / 0.61\\
    \midrule
    Implication & \makecell[l]{diabetes\\mellitus} & \makecell[l]{... Medications on Admission: lipitor 40mg\\po qday \textbf{\textcolor{red}{metformin}}\tnote{4} 1000mg po bid ...} & 11 / 0.66 & 2 / 0.86\\
    \bottomrule
  \end{tabular}
\begin{tablenotes}  
    \footnotesize  
    \item[1] Dilantin is a brand name of phenytoin.
    \item[2] HTN is the common abbreivation for hypertension.
    \item[3] IVC filter is a subtype of interruption of the vena cava.
    \item[4] Metformin is a common hypolycemic agent.
  \end{tablenotes}  
\end{threeparttable}
\end{table*}

\end{document}